\documentclass[aps,preprint,prd,showpacs,nofootinbib,superscriptaddress]{revtex4}

\usepackage{mathrsfs}
\usepackage{latexsym}
\usepackage{amsmath}
\usepackage{graphicx}
\usepackage{subfigure}
\usepackage{dcolumn}
\usepackage{bm}
\usepackage{amssymb}
\usepackage{latexsym}
\usepackage{ulem}
\usepackage{comment}
\usepackage{color}
\usepackage[colorlinks,linkcolor=magenta,anchorcolor=cyan,citecolor=blue]{hyperref}

\def\be{\begin{equation}}
\def\ee{\end{equation}}
\def\ba{\begin{eqnarray}}
\def\ea{\end{eqnarray}}

\begin{document}

\begin{flushleft}
{\footnotesize

}
\end{flushleft}

\title{Testing the wormhole echo hypothesis for GW231123}

\author{Qi Lai}
\email{laiqi23@mails.ucas.ac.cn}
\affiliation{School of Fundamental Physics and Mathematical Sciences, Hangzhou Institute for Advanced Study, UCAS, Hangzhou 310024, China}
\affiliation{Institute of Theoretical Physics, Chinese Academy of Sciences, P.O. Box 2735, Beijing 100190, China}
\affiliation{University of Chinese Academy of Sciences, Beijing 100190, China}

\author{Qing-Yu Lan}
\email{lanqingyu19@mails.ucas.ac.cn} \affiliation{School of
Physical Sciences, University of Chinese Academy of Sciences,
Beijing 100049, China}

\author{Zhan-He Wang}
\email{wangzhanhe19@mails.ucas.ac.cn} \affiliation{School of
Physical Sciences, University of Chinese Academy of Sciences,
Beijing 100049, China}

\author{Yun-Song Piao}
\email{yspiao@ucas.ac.cn}
\affiliation{School of Fundamental Physics
and Mathematical Sciences, Hangzhou Institute for Advanced Study,
UCAS, Hangzhou 310024, China}
\affiliation{Institute of
Theoretical Physics, Chinese Academy of Sciences, P.O. Box 2735,
Beijing 100190, China}
\affiliation{School of Physical
Sciences, University of Chinese Academy of Sciences, Beijing
100049, China}
\affiliation{International Centre for Theoretical
Physics Asia-Pacific, University of Chinese Academy of Sciences,
100190 Beijing, China}

\begin{abstract}
The short-duration gravitational-wave (GW) event GW231123 has component masses inferred to lie in the pair-instability mass gap and exhibits a burst-like morphology without a clear inspiral phase, making it an interesting target for testing alternatives to the standard binary black hole (BBH) interpretation. Motivated by its phenomenological similarity to GW190521, we test a wormhole-echo interpretation by modeling a leading echo pulse with a phenomenological \texttt{sine-Gaussian} wavepacket. Bayesian model comparison against a BBH interpretation described by the \texttt{IMRPhenomXPHM-SpinTaylor} waveform yields $\ln \mathcal{B}^{\rm Echo}_{\rm BBH}=1.87$. This preference is sensitive to the BBH waveform family and Echo-prior choices, and the observed statistic is not unusual within the targeted BBH-null ensembles. The positive Bayes factor therefore does not provide robust evidence for a wormhole-echo interpretation, with the analyses considered here supporting the standard BBH interpretation of GW231123.
\end{abstract}

\maketitle

\section{Introduction}
\label{sec:introduction}

The detection of gravitational waves (GWs) has enabled stringent
tests of general relativity and the study of compact-object
populations. To date, most events reported by the
LIGO--Virgo--KAGRA (LVK) Collaboration follow the standard
inspiral--merger--ringdown (IMR) morphology expected from binary
black hole (BBH) or binary neutron star (BNS)
coalescences~\cite{LIGOScientific:2018mvr,LIGOScientific:2020ibl,LIGOScientific:2021usb,KAGRA:2021vkt,LIGOScientific:2025slb}.
However, the observation of short-duration events lacking a clear
inspiral phase poses potential challenges to this standard
paradigm.

The event GW231123, detected on November 23, 2023, is notable for
its extremely short duration of $\sim 10$\,ms and the absence of a
distinct inspiral phase in the sensitive 30--80\,Hz band. Under
the standard BBH interpretation, the LVK Collaboration reports
component masses of $137^{+22}_{-17}~\rm M_{\odot}$ and
$103^{+20}_{-52}~\rm M_{\odot}$ (90\% credible level), making this
system the most massive observed to
date~\cite{LIGOScientific:2025rsn}. The inferred component masses lie within or above the
pair-instability mass gap, which is difficult to accommodate
within standard stellar-evolution
expectations~\cite{Woosley:2016hmi,Hendriks:2023yrw,Farmer:2019jed,Farmer:2020xne}.
While the LVK Collaboration favors a BBH origin, the signal's
brevity and extreme parameters have sparked debate, prompting
consideration of alternative formation
channels~\cite{Li:2025fnf,Delfavero:2025lup,Li:2025pyo,Kiroglu:2025vqy,Bartos:2025pkv,Tanikawa:2025fxw,Croon:2025gol,Popa:2025dpz,Stegmann:2025cja,Gottlieb:2025ugy}
or exotic
origins~\cite{Yuan:2025avq,Nojiri:2025heh,DeLuca:2025fln,Shan:2025jpt,Cuceu:2025fzi}.

Wormholes are hypothetical spacetime
structures connecting distant regions or, in some constructions, different universes~\cite{Morris:1988cz,Morris:1988tu,Visser:1989kh}, and
have been discussed as potential arenas for quantum-gravity
effects and as a possible ingredient in resolving the black hole
information problem. In the GW context, post-merger ``echoes'' have been proposed as possible signatures of horizonless ultracompact remnants~\cite{Cardoso:2016rao,Cardoso:2016oxy,Rahman:2021kwb} or of modifications to the effective potential and near-horizon structure~\cite{Dong:2020odp,Guo:2022umh,Cardoso:2019apo,Coates:2021dlg}. Studies have shown that if the post-merger remnant of a BBH merger were a
wormhole, the post-merger response could partially enter the
throat and undergo repeated reflections between the photon-sphere
potential barriers on either side, producing a sequence of
pulse-like signals (echoes), e.g.~\cite{Mark:2017dnq,Cardoso:2017cqb,Bueno:2017hyj,Wang:2018mlp,GalvezGhersi:2019lag,Bronnikov:2019sbx,Maggio:2019zyv,Li:2019kwa,Liu:2020qia,Yang:2024prm,Siemonsen:2024snb}. Searches for gravitational-wave echoes remain an active area of investigation~\cite{Abedi:2016hgu,Westerweck:2017hus,Lo:2018sep,Nielsen:2018lkf,Uchikata:2019frs,Wang:2020ayy,LIGOScientific:2020tif,Ren:2021xbe,Abedi:2021tti,Abedi:2022bph,Wu:2023wfv,Uchikata:2023zcu,Wu:2025enn}. A leading echo pulse may appear as a short-duration burst-like signal.

GW231123 shares several morphological characteristics with
GW190521, another short-duration event that posed a challenge to standard template-bank analyses and whose inferred component masses
also lie in the pair-instability mass gap~\cite{LIGOScientific:2020iuh}.
In our previous study, we tested an interpretation in terms of an isolated leading echo pulse for GW190521~\cite{Lai:2025skp} and found that the
BBH model was favored under that analysis, with
$\ln \mathcal{B}^{\rm Echo}_{\rm BBH}\approx -2.9$.
Here, we apply the same model-comparison framework to GW231123.
Under the fiducial configuration, the analysis yields a modest
positive relative preference for the Echo model, with
$\ln\mathcal{B}^{\rm Echo}_{\rm BBH}=1.87$.
We then assess the sensitivity of this result to the adopted BBH
waveform, Echo priors, and noise realization. The fiducial positive
Bayes factor is sensitive to waveform and prior choices and is not
statistically unusual within the targeted BBH-null ensembles.
It therefore does not provide a robust basis for a wormhole-echo
interpretation of GW231123.

\section{Methods and analysis setup}
\label{sec:setup}

\subsection{Waveform template}
\label{sec:waveform_template}

In our previous work on GW190521~\cite{Lai:2025skp}, we tested the
hypothesis that the event could be associated with a wormhole
remnant in another universe, whose post-merger response manifests
as an echo-like signal in our universe. In this picture, a
short-lived wormhole may form as an intermediate state after a BBH
merger and then rapidly pinch off, so that only the leading pulse
entering our universe is expected to be observable
\cite{Wang:2018mlp}. More generally, if the wormhole pinches off
more slowly and produces a sequence of echoes, successive pulses
are expected to be filtered and attenuated by the photon-sphere
barrier and finite effective reflectivity, potentially falling
below the sensitivity of current detectors
\cite{Maggio:2019zyv,Abedi:2021tti}.

Motivated by the possibility that only a leading pulse is observable, we model the observed signal as a phenomenological \texttt{sine-Gaussian} wavepacket\footnote{The wormhole echo waveforms investigated in
Ref.~\cite{Bueno:2017hyj} motivate a phenomenological
\texttt{sine-Gaussian} description. A similar modeling approach  
was also adopted in Ref.~\cite{Lo:2018sep}.}: 
\be
h(t)=A_{m}\exp\!\left[-\frac{(t-t_{c})^{2}}{2\beta^{2}}\right]\cos\!\left[2\pi
f_{n} (t-t_{c})+\varphi\right], \label{Waveformecho} 
\ee 
where
$f_{n}$ (in Hz) is the central frequency, $\beta$ is the
characteristic width, and $\varphi$ is the reference phase.
Throughout this work, time is defined relative to the trigger time
(GPS $1384782888.634$), and $t_c$ denotes the central time of the
pulse at the geocenter. The strain measured by each detector is
obtained by projecting the two polarizations onto the detector
response using the antenna-pattern functions. We further
parameterize the overall amplitude by a normalized factor $A_m =
\mathcal{A}/A_{\rm ref}$, where $\mathcal{A}$ is the physical
strain amplitude and $A_{\rm ref}$ is a reference strain. In this
work, we fix $A_{\rm ref}=10^{-19}$ at a reference distance of
$1\,\mathrm{Mpc}$. This wavepacket is intended as a model-agnostic
description of a leading echo pulse and does not explicitly
incorporate spin-induced modifications to the effective potential
or the echo spectrum. Given the theoretical uncertainty in the
detailed echo morphology, we consider $(f_n,\beta)$ as effective
parameters that absorb part of such modeling
uncertainty~\cite{Bueno:2017hyj,Maselli:2017tfq}. Neglecting spin
effects is therefore a limitation of the present analysis, which
we leave to future work with more physically motivated
spinning-echo models.

For the BBH hypothesis, the LVK Collaboration analyzed GW231123
with multiple waveform approximations and reported significant
parameter-estimation variations across models (see Table~3 of
Ref.~\cite{LIGOScientific:2025rsn}). To facilitate a direct comparison with our previous GW190521 study~\cite{Lai:2025skp}, we use \texttt{IMRPhenomXPHM-SpinTaylor}~\cite{Garcia-Quiros:2020qpx,Colleoni:2024knd} as the fiducial BBH baseline. This approximation extends
\texttt{IMRPhenomXPHM} by incorporating generic spin precession
through a time-domain prescription, improving the phase evolution
for binaries with strong spin-induced orbital-plane precession.
Given the LVK inference of large spin effects and a precession
parameter $\chi_p \approx 0.77$ for
GW231123~\cite{LIGOScientific:2025rsn},
\texttt{IMRPhenomXPHM-SpinTaylor} provides a suitable baseline for
our BBH-versus-echo model comparison.

\subsection{Bayesian inference and model selection}
\label{subsec:bayes_setup}

For the echo-for-wormhole waveform, we adopt the same prior ranges
as in our previous analysis~\cite{Lai:2025skp}, with two
modifications. First, since GW231123 has a larger strain amplitude
than GW190521, we extend the upper bound of the normalized
amplitude to $A_m^{\max}=0.3$ so that the posterior support
remains well contained within the sampling domain. Second, because
no electromagnetic counterpart or other external constraint on the
sky location has been identified~\cite{He:2025osl}, we adopt an
uninformative isotropic prior for the sky position; this differs
from our GW190521 analysis, where a candidate counterpart enabled
an informative sky-location prior~\cite{Graham:2020gwr}. The prior
choices for the remaining intrinsic and extrinsic parameters are
unchanged and are summarized in Table~\ref{Prior_echo}. Note that,
because the \texttt{sine-Gaussian} template is parameterized by a
normalized amplitude rather than an explicit luminosity distance,
we do not perform distance marginalization as in the LVK BBH
analysis. 

\begin{table}[t]
\caption{\label{Prior_echo}
Prior ranges for the echo-for-wormhole waveform parameters.
The amplitude $A_m$ is log-uniform;
$(\beta,f_n,t_c,\psi,\varphi)$ are uniform in the intervals shown. The inclination and sky
position are isotropic, corresponding to uniform priors in
$\cos\iota$, $\alpha$, and $\sin\delta$, respectively.}
\centering
\begin{ruledtabular}
\begin{tabular}{lc}
Parameter & Prior range \\
\hline
Amplitude, $A_m$
    & $[10^{-4},\,0.3]$ \\
Echo width, $\beta$
    & $[0.0058,\,0.026]~\mathrm{s}$ \\
Central frequency, $f_n$
    & $[52,\,70]~\mathrm{Hz}$ \\
Polarization angle, $\psi$
    & $[0,\,\pi]$ \\
Phase, $\varphi$
    & $[0,\,2\pi]$ \\
Inclination, $\iota$
    & $[0,\,\pi]$ \\
Right ascension, $\alpha$
    & $[0,\,2\pi]$ \\
Declination, $\delta$
    & $[-\pi/2,\,\pi/2]$ \\
Geocenter time, $t_c$
    & $[1384782888.5,\,1384782888.7]~\mathrm{s}$ \\
\end{tabular}
\end{ruledtabular}
\end{table}

The strain data of GW231123 from the Hanford (H1) and Livingston
(L1) detectors are publicly available through the Gravitational Wave Open Science Center (GWOSC)~\cite{LIGOScientific:2025snk}. We analyze a
$4\,\mathrm{s}$ segment centered on the trigger time (GPS
$1384782888.634$), sampled at $512\,\mathrm{Hz}$, and perform the
analysis over the frequency band $[20,\,256]\,\mathrm{Hz}$.
Parameter estimation is carried out with \texttt{Parallel
Bilby}~\cite{Ashton:2018jfp,Smith:2019ucc} using the nested
sampler \texttt{Dynesty}~\cite{Speagle:2020,Skilling:2006gxv}. The
detector-noise power spectral densities (PSDs) are taken to be the
off-source PSD estimates generated within the \texttt{Bilby}
workflow and are held fixed during sampling. We use $N_{\rm
live}=1500$ live points and terminate the sampler when
$\Delta\ln\mathcal{Z}=0.1$. Throughout, we set the reference
frequency to $f_{\rm ref}=10\,\mathrm{Hz}$.

To investigate the possible origin of GW231123, we compare the
echo-for-wormhole hypothesis with the BBH hypothesis by computing
Bayes factors. For a given hypothesis $\mathcal{H}_i$, the Bayes
factor relative to the noise-only hypothesis $\mathcal{H}_0$ is
defined as 
\be
\mathcal{B}_i =
\frac{P(d|\mathcal{H}_{i})}{P(d|\mathcal{H}_{0})},
\label{BayesFactor} 
\ee
where $P(d|\mathcal{H}_{i})$ denotes the
Bayesian evidence under $\mathcal{H}_i$ and $P(d|\mathcal{H}_{0})$
corresponds to the noise-only scenario. As our model-selection
statistic, we use the log Bayes factor between the two signal
hypotheses, 
\be
\ln \mathcal{B}^{\rm Echo}_{\rm BBH} =
\ln\!\left[\frac{P(d|\mathcal{H}_{\rm Echo})}{P(d|\mathcal{H}_{\rm
BBH})}\right].\label{Eq:Bayes_factor} 
\ee
Positive values indicate relative preference for the echo-for-wormhole hypothesis over the BBH hypothesis under the specified waveform implementations and priors, while negative values favor the BBH hypothesis.

However, a positive model preference does not by itself indicate
whether the observed value is unusual under the BBH hypothesis; see, e.g., Refs.~\cite{Nielsen:2018lkf,Ren:2021xbe}.
We define
$T_{\rm obs}=\ln \mathcal{B}^{\rm Echo}_{\rm BBH}$ as the
model-selection statistic measured from the on-source data. Given a targeted
BBH-null calibration ensemble $\Lambda$, the corresponding empirical $p$-value
is defined as
\be
p_\Lambda(T_{\rm obs}) =
P\!\left(T\geq T_{\rm obs}\mid
\mathcal{H}_{\rm BBH},\Lambda\right)
=
\int_{T_{\rm obs}}^\infty
f_\Lambda(T|\mathcal{H}_{\rm BBH})\,dT,
\label{Eq:BBH_null_pvalue}
\ee
where $f_\Lambda$ denotes the sampling distribution of $T$ under the specified
BBH-null calibration configuration. This $p$-value measures the fraction of BBH-null realizations in the specified ensemble that yield a model-selection statistic no smaller than the on-source value. In the following BBH-null calibration, we estimate this quantity from an ensemble constructed by injecting a GW231123-like BBH signal into independent noise realizations.

\section{Results}
\label{sec:result}

\subsection{Parameter inference}
\label{subsec:parameter_inference}

\begin{figure*}[tbp]
\centering
\includegraphics[width=0.8\textwidth]{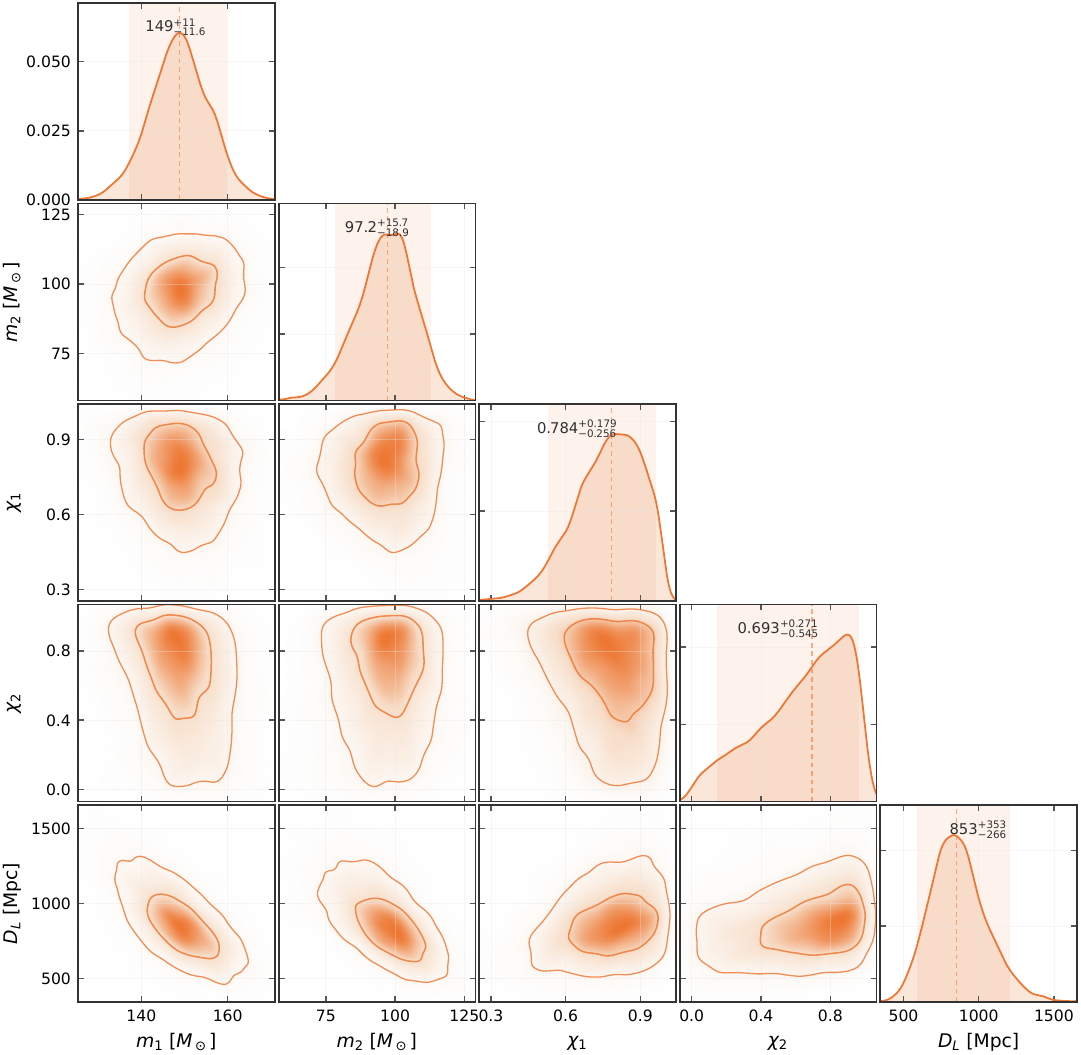}
\caption{\label{IMR_corner}
Posterior distributions for selected BBH parameters
(source-frame component masses, dimensionless spin magnitudes and
luminosity distance) inferred under the BBH
hypothesis using \texttt{IMRPhenomXPHM-SpinTaylor}. The vertical
lines indicate the medians, and the shaded regions show the 90\%
credible intervals.}
\end{figure*}

Under the BBH hypothesis, we use \texttt{IMRPhenomXPHM-SpinTaylor}
for Bayesian parameter estimation. Figure~\ref{IMR_corner} shows posterior distributions for selected
BBH parameters. We infer component masses of
$m_1=149^{+11}_{-12}M_\odot$ and
$m_2=97^{+16}_{-19}M_\odot$, and dimensionless spin
magnitudes $\chi_1=0.78^{+0.18}_{-0.26},$ and
$\chi_2=0.69^{+0.27}_{-0.55}.$ (median and 90\% credible
intervals). These estimates are consistent with the values
reported by the LVK Collaboration obtained with the same waveform
approximation~\cite{LIGOScientific:2025rsn}, supporting the
internal consistency of our analysis setup. Consistent with the
LVK findings, the inferred masses place the primary component
$m_1$ above and the secondary $m_2$ within the pair-instability
mass gap.

\begin{figure*}[tbp]
\centering
\includegraphics[width=0.95\textwidth]{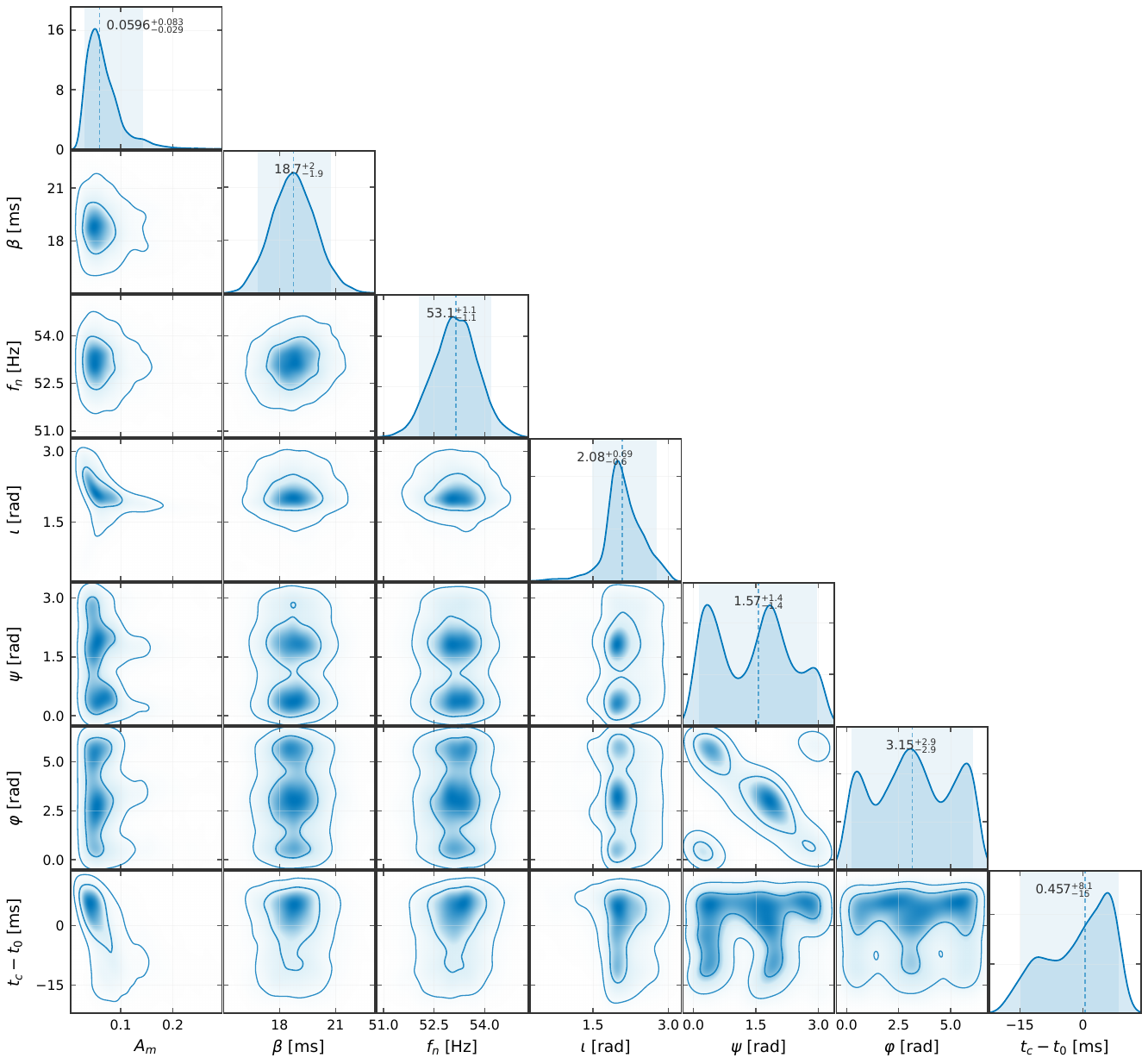}
\caption{\label{Echo_corner}
Posterior distributions of the echo-for-wormhole
waveform parameters inferred with the phenomenological
\texttt{sine-Gaussian} wavepacket template. The vertical lines
indicate the medians and the shaded regions show the 90\% credible
intervals.}
\end{figure*}

We next
perform Bayesian parameter estimation with the echo-for-wormhole
model and show the corresponding posterior distributions in
Fig.~\ref{Echo_corner}. The posteriors indicate that the data
constrain the key parameters of the phenomenological echo
template. In particular, the central frequency is inferred to be
$f_{n}=53.1^{+1.1}_{-1.1}\,\mathrm{Hz}$, which lies in the frequency
range relevant for the leading post-merger response in
representative wormhole-echo models~\cite{Bueno:2017hyj}. Within
our \texttt{sine-Gaussian} parametrization, the tight constraint
on $f_n$ implies that the preferred waveform power is concentrated
in a narrow band in the sensitive region. In the wormhole-echo
picture, such a narrow-band response is consistent with the
expectation that the photon-sphere potential barrier acts as a
frequency-dependent filter, so that only a limited range of
frequencies can efficiently penetrate and contribute to an
observable leading echo pulse~\cite{GalvezGhersi:2019lag}.

\subsection{Fiducial model comparison}
\label{subsec:model_comparison}

\begin{figure*}[tbp]
\centering
\includegraphics[width=0.45\textwidth]{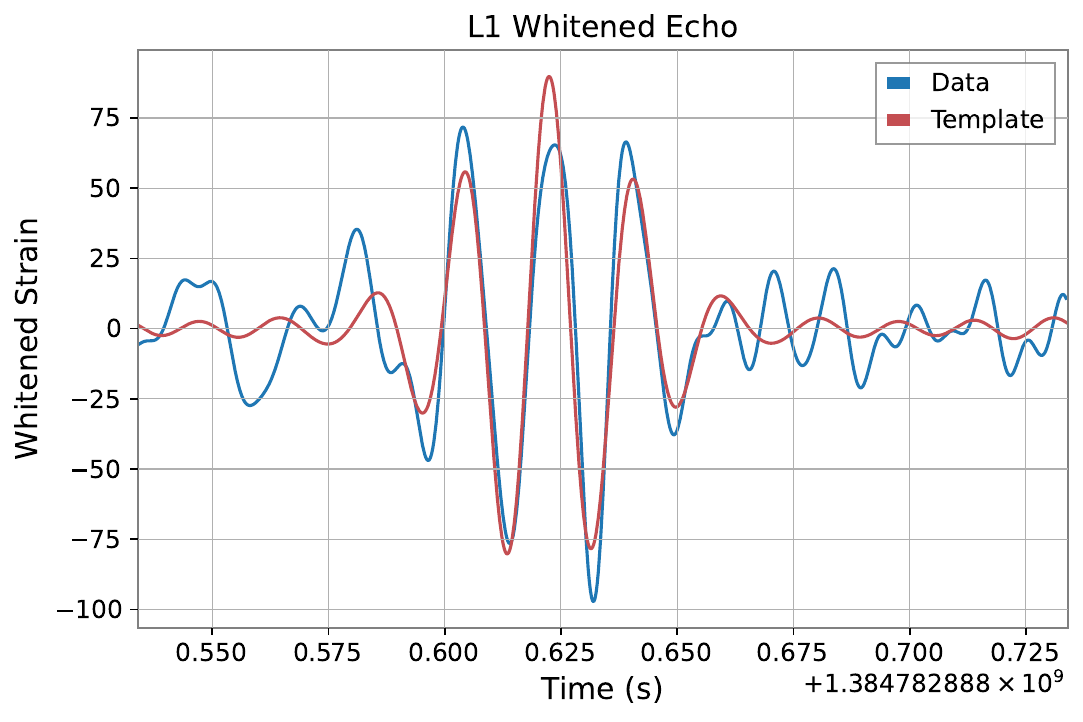}
\includegraphics[width=0.45\textwidth]{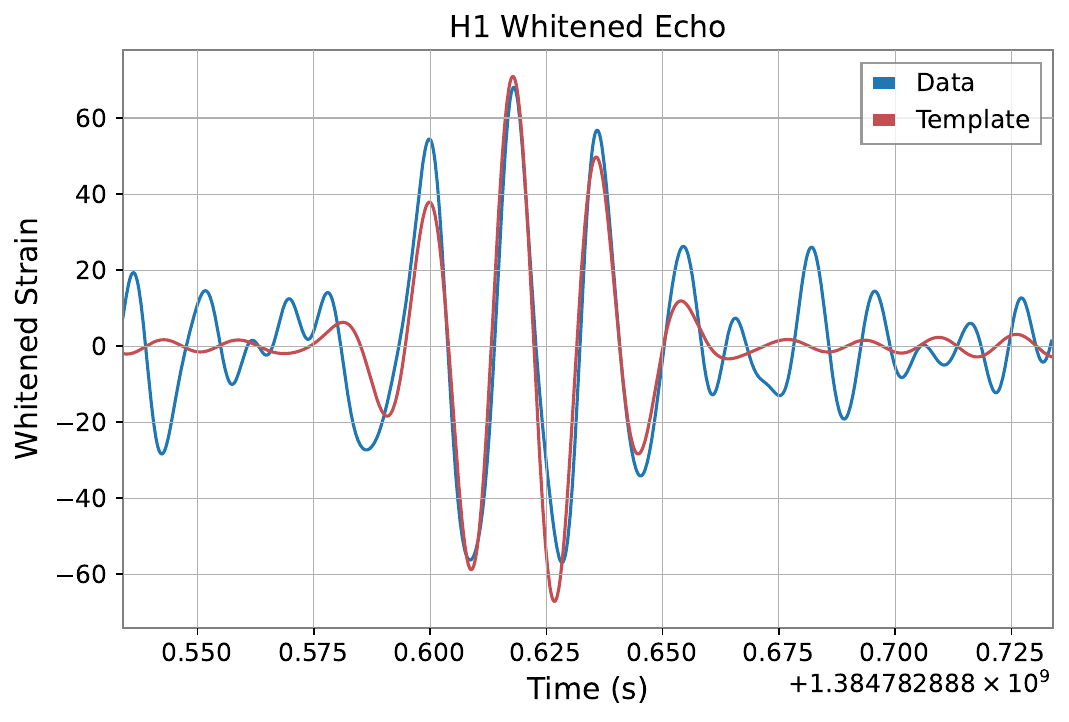}
\includegraphics[width=0.45\textwidth]{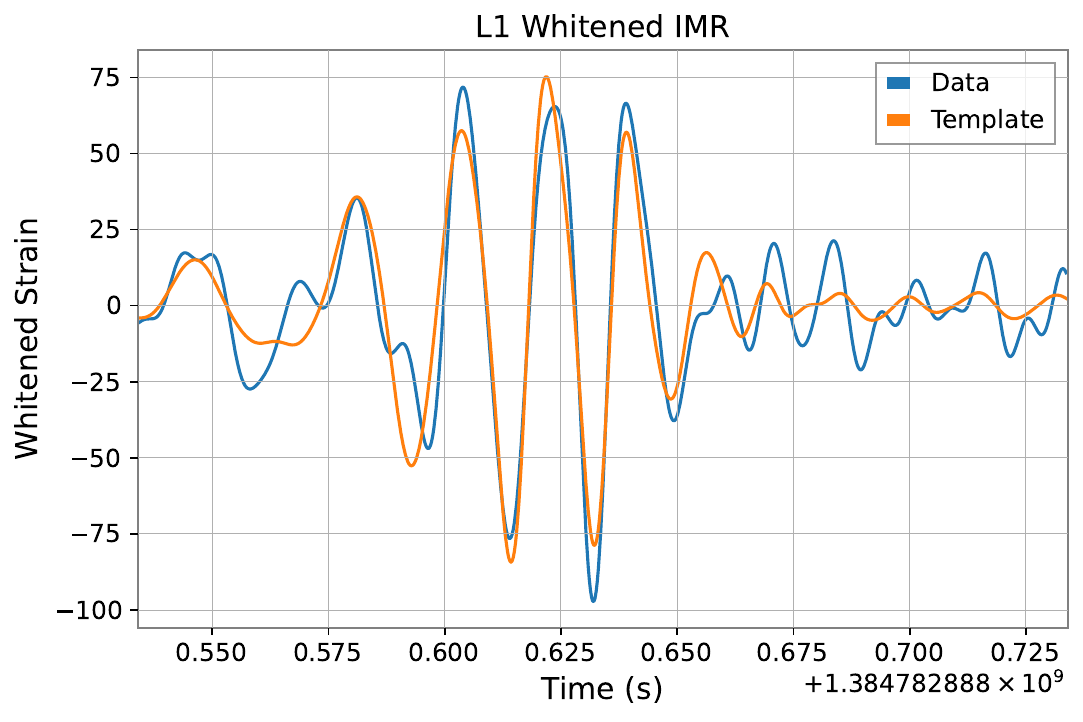}
\includegraphics[width=0.45\textwidth]{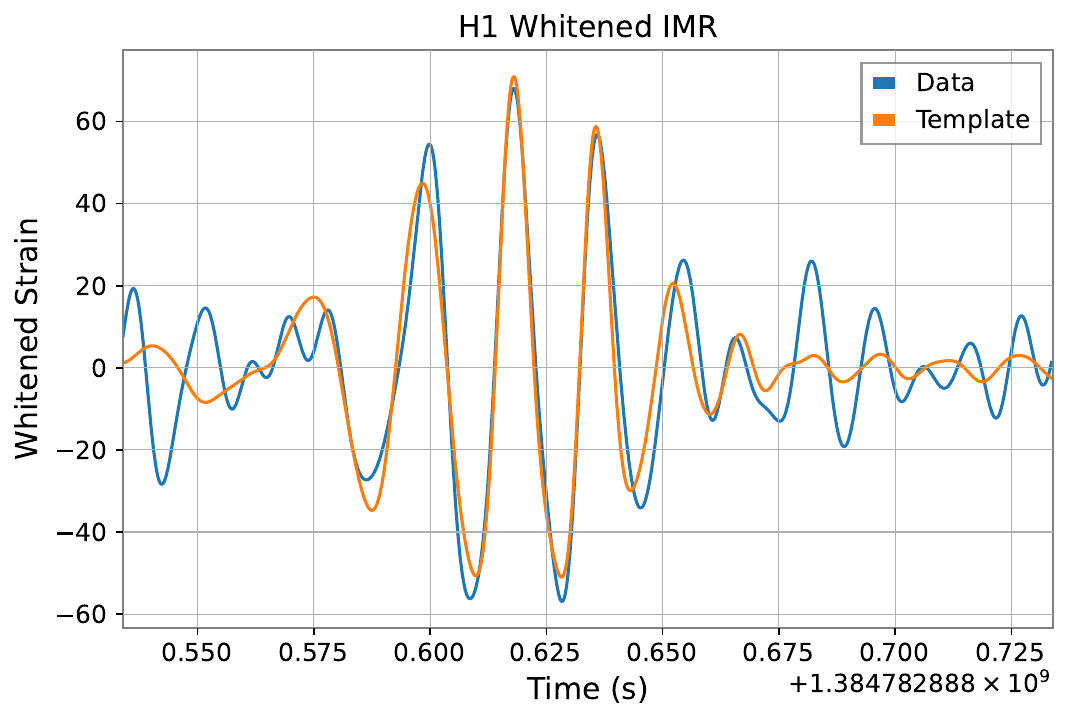}
\caption{\label{Waveform}
Whitened strain data and maximum-likelihood waveform
reconstructions for the L1 (left) and H1 (right) detectors around
the trigger time (GPS $1384782888.634$). The top row shows the
echo-pulse model, while the bottom row shows the BBH model using
the \texttt{IMRPhenomXPHM-SpinTaylor} approximant. For
visualization, the data and waveform reconstructions are whitened
using the off-source noise PSDs and band-pass filtered over
$20$--$128~\mathrm{Hz}$; the whitening procedure is defined in
Eq.~\eqref{Eq:whiten}.}
\end{figure*}

To visualize the signal relative to the frequency-dependent
detector noise, we whiten the strain data in the frequency domain
using the off-source noise PSD $S_n(f)$, \be
\tilde{h}_{w}(f)=\frac{\tilde{h}(f)}{\sqrt{S_n(f)}} ,
\label{Eq:whiten} \ee where $\tilde{h}(f)$ denotes the Fourier
transform of the detector strain. This procedure normalizes each
frequency bin by the local noise amplitude and facilitates
qualitative time-domain comparisons of the signal morphology. Fig.~\ref{Waveform} shows the whitened strain data together with
the maximum-likelihood waveform reconstructions under the two
hypotheses. For visualization only, we apply a band-pass filter
over $20$--$128~\mathrm{Hz}$ to suppress high-frequency noise and
improve the clarity of the time-domain features; the Bayesian
analysis itself is performed over $20$--$256~\mathrm{Hz}$ as
described in Sec.~\ref{subsec:bayes_setup}. Both templates
reproduce the ${\sim}5$ visible oscillation cycles near the peak
amplitude, indicating that each model captures the dominant
features of the observed event.

Using the posterior samples, we compute the matched-filter
signal-to-noise ratios (SNRs) in H1 and L1 for each hypothesis,
and quote the corresponding network SNRs in Table~\ref{SNR_result}.
Both models yield $\mathrm{SNR}_{\rm net}\gtrsim 8$, indicating
that the data contain a coherent GW event rather than being
consistent with the noise-only hypothesis. Notably, the echo and
BBH hypotheses recover comparable network SNRs, $\mathrm{SNR}_{\rm
net}\simeq 20.99$ and $21.93$, respectively, consistent with the
LVK estimate $\mathrm{SNR}_{\rm net}=22.4^{+0.2}_{-0.3}$ for the
BBH interpretation~\cite{LIGOScientific:2025rsn}. This similarity
illustrates that SNR alone is not sufficient to discriminate
between the two signal hypotheses, motivating a Bayesian model
comparison.

\begin{table}[t]
\caption{\label{SNR_result}
Matched-filter SNRs recovered under the echo-for-wormhole and BBH
hypotheses in the H1 and L1 detectors, together with the
corresponding network values.}
\centering
\begin{ruledtabular}
\begin{tabular}{lcc}
SNR & Echo & BBH \\
\hline
$\mathrm{SNR}_{\rm H1}$
    & $13.05$ & $13.46$ \\
$\mathrm{SNR}_{\rm L1}$
    & $16.44$ & $17.32$ \\
$\mathrm{SNR}_{\rm net}$
    & $20.99$ & $21.93$ \\
\end{tabular}
\end{ruledtabular}
\end{table}

We quantify the relative support for the echo-for-wormhole versus
BBH interpretations using the log Bayes factor defined in
Eq.~\eqref{Eq:Bayes_factor}. For GW231123, we find \be \ln
\mathcal{B}^{\rm Echo}_{\rm BBH}=1.87, \ee 
which indicates a modest positive relative preference for our
echo-for-wormhole model over the BBH model under the fiducial
\texttt{IMRPhenomXPHM-SpinTaylor} waveform and the Echo priors
listed in Table~\ref{Prior_echo}. This result should not be
interpreted as a detection-significance measure.

Although the fiducial analysis yields a positive relative
preference for our echo-for-wormhole model, this result is obtained under a
specific choice of BBH waveform and Echo priors and for the
observed detector-noise realization. Given the short signal
duration and limited number of in-band cycles, we therefore assess
the robustness of this Bayesian preference by examining its
dependence on analysis choices and by calibrating the observed
model-selection statistic against targeted BBH-null ensembles.

\section{Robustness and empirical calibration}
\label{sec:robustness}

\subsection{BBH waveform systematics}
\label{subsec:bbh_waveform_systematics}

In the fiducial analysis, we adopted \texttt{IMRPhenomXPHM-SpinTaylor} 
(XPHM) as the BBH waveform template. Given the high masses and spins of GW231123, together with the
limited number of in-band cycles, parameter inference and
model-selection conclusions for this event can be particularly
sensitive to waveform systematics, as shown in
Refs.~\cite{LIGOScientific:2025rsn,LIGOScientific:2026fcf}; see also Refs.~\cite{Chan:2025kyu,Bini:2026kwz,Jan:2025zcm}. 
We therefore repeat the BBH analysis using different waveform
approximants, including \texttt{IMRPhenomXO4a} (XO4a)~\cite{Thompson:2023ase},
\texttt{NRSur7dq4} (NRSur)~\cite{Varma:2019csw}, and \texttt{SEOBNRv5PHM} (v5PHM)~\cite{Ramos-Buades:2023ehm}, while keeping all other
analysis choices unchanged, including the Echo sampling adopted
in the fiducial analysis. This analysis examines how the choice of BBH waveform affects the Bayesian model comparison and, consequently, assesses the robustness of the inferred $\ln \mathcal{B}^{\rm Echo}_{\rm BBH}$ with respect to BBH waveform systematics.

\begin{figure*}[tbp]
\centering
\includegraphics[width=0.8\textwidth]{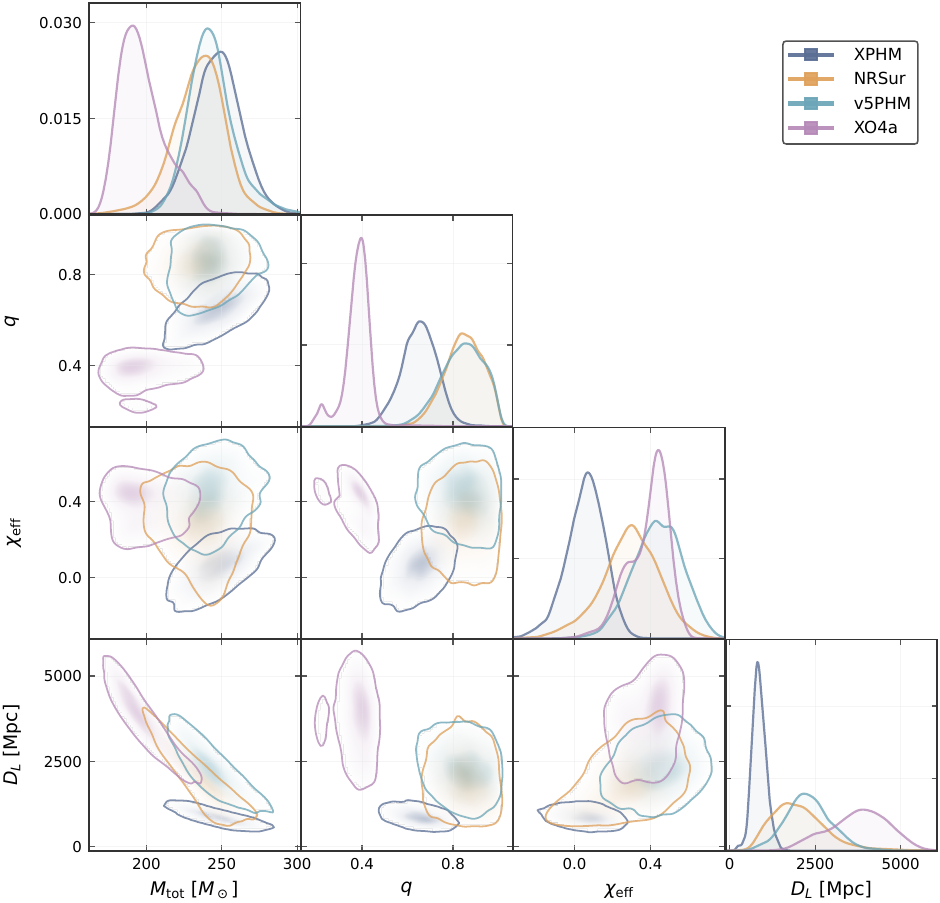}
\caption{\label{Waveform_systematics_corner}
Posterior distributions of the total mass $M_{\rm tot}$, mass ratio $q$,
effective spin $\chi_{\rm eff}$, and luminosity distance $D_L$ obtained using
four BBH waveform models. The diagonal panels show the marginalized posterior
densities, while the off-diagonal panels show the joint posterior distributions
with 90\% credible regions.}
\end{figure*}

Figure~\ref{Waveform_systematics_corner} shows posterior distributions that are broadly consistent with the LVK Collaboration's inference results obtained with different BBH waveform models~\cite{LIGOScientific:2025rsn}. The XPHM, NRSur, and v5PHM results overlap in a high-$M_{\rm tot}$ region, whereas XO4a favors a substantially lower total mass. The posterior distributions of $q$, $\chi_{\rm eff}$, and $D_L$ also exhibit systematic differences: XO4a favors a lower mass ratio, a more positive effective spin, and a larger luminosity distance, consistent with the waveform-dependent trends reported in
Ref.~\cite{Bini:2026kwz}, while NRSur and v5PHM exhibit distinct correlations among these
parameters. Therefore, for GW231123, changes in the adopted BBH waveform model shift the inferred binary configuration rather than merely broadening a similar posterior distribution.

The Echo-versus-BBH Bayes factors are summarized in
Table~\ref{Waveform_systematics_BF}. The fiducial XPHM analysis gives
$\ln \mathcal{B}^{\rm Echo}_{\rm BBH}=1.87$, whereas XO4a, NRSur, and v5PHM
give $-8.01$, $-3.64$, and $-6.30$, respectively. Since the Echo model and its
sampling procedure are fixed across these analyses, the negative values
indicate that the BBH hypothesis is favored by the Bayesian evidence under
these waveform choices. The significant variation in
$\ln \mathcal{B}^{\rm Echo}_{\rm BBH}$ further demonstrates that the impact of
waveform systematics extends beyond parameter inference and affects the
Bayesian model comparison. These results do not determine which BBH waveform
model provides the most accurate description of the signal, but show that the
fiducial value of $1.87$ is conditional on the adopted BBH waveform model and
that the corresponding Echo preference is not robust against BBH waveform
systematics.

\begin{table}[t]
\caption{\label{Waveform_systematics_BF}
Echo-versus-BBH log Bayes factors obtained using different BBH
waveform models. The \texttt{IMRPhenomXPHM-SpinTaylor} result
corresponds to the fiducial analysis; all other analysis settings
are kept unchanged when varying the BBH waveform model.}
\centering
\begin{ruledtabular}
\begin{tabular}{lc}
BBH waveform model
    & $\ln\mathcal{B}^{\rm Echo}_{\rm BBH}$ \\
\hline
\texttt{IMRPhenomXPHM-SpinTaylor}
    & $1.87$ \\
\texttt{IMRPhenomXO4a}
    & $-8.01$ \\
\texttt{NRSur7dq4}
    & $-3.64$ \\
\texttt{SEOBNRv5PHM}
    & $-6.30$ \\
\end{tabular}
\end{ruledtabular}
\end{table}

\subsection{Echo-prior sensitivity}
\label{subsec:echo_prior_sensitivity}

In the fiducial analysis, we adopt the prior ranges for the three
Echo feature parameters listed in Table~\ref{Prior_echo}. Since the
prior ranges of these parameters are not fixed by an established
physical prescription for the present phenomenological
echo-for-wormhole model, their prior choices are not uniquely
determined~\cite{Maselli:2017tfq}. Previous Bayesian echo analyses
have also emphasized the importance of the adopted prior
specification, particularly when some Echo parameters are only
weakly constrained~\cite{Nielsen:2018lkf}. We therefore investigate the dependence of the inferred results on representative prior alternatives by considering narrower and wider prior ranges for the three Echo feature parameters $A_m$, $f_n$, and $\beta$, while keeping the remaining analysis choices unchanged, as summarized in Table~\ref{Echo_prior_sensitivity_table}. We also consider a joint-wide configuration in which the prior ranges of all three Echo feature parameters are simultaneously extended.

\begin{figure*}[tbp]
\centering
\includegraphics[width=0.7\textwidth]{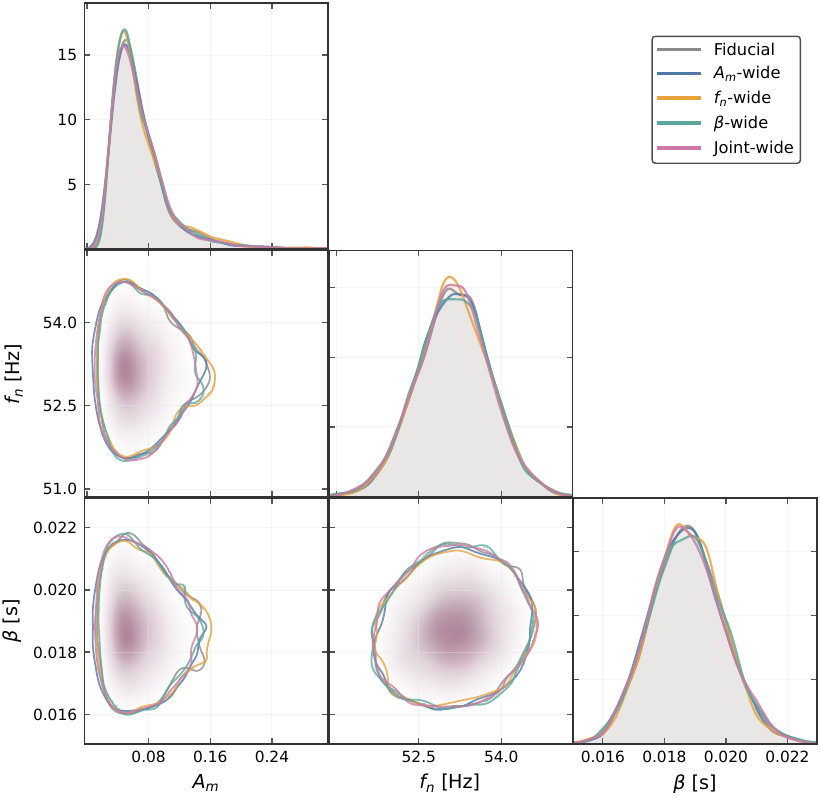}
\caption{\label{Echo_prior_posterior_corner}
Posterior distributions of the Echo feature parameters 
$A_m$, $f_n$, and $\beta$ for the fiducial prior, $A_m$-wide, $f_n$-wide, 
$\beta$-wide, and joint-wide prior configurations. The diagonal panels show 
marginalized posterior densities, and the off-diagonal contours enclose 90\% 
credible regions.}
\end{figure*}

Figure~\ref{Echo_prior_posterior_corner} shows substantial overlap
among the high-posterior-density regions obtained with different
prior configurations, with the inferred parameter regions and
main correlations remaining stable under the wider priors. These posterior comparisons
alone, however, do not fully characterize the Bayesian evidence, which depends
on the prior-weighted likelihood integral.

\begingroup
\squeezetable
\begin{table*}[t]
\caption{\label{Echo_prior_sensitivity_table}
Prior specifications and resulting values of
$\ln\mathcal{B}^{\rm Echo}_{\rm BBH}$ for the tested Echo-prior
configurations. Here, $\mathrm{U}(a,b)$ and
$\log\mathrm{U}(a,b)$ denote uniform and log-uniform priors,
respectively.}
\centering
\begin{ruledtabular}
\begin{tabular}{lcccc}
Prior configuration
& $A_m$ prior
& $f_n$ prior ($\mathrm{Hz}$)
& $\beta$ prior ($\mathrm{s}$)
& $\ln\mathcal{B}^{\rm Echo}_{\rm BBH}$ \\
\hline
Fiducial
& $\log\mathrm{U}(10^{-4},0.3)$
& $\mathrm{U}(52,70)$
& $\mathrm{U}(0.0058,0.026)$
& $1.87$ \\

$A_m$-narrow
& $\log\mathrm{U}(10^{-4},0.175)$
& $\mathrm{U}(52,70)$
& $\mathrm{U}(0.0058,0.026)$
& $2.21$ \\

$A_m$-wide
& $\log\mathrm{U}(10^{-5},1.0)$
& $\mathrm{U}(52,70)$
& $\mathrm{U}(0.0058,0.026)$
& $0.71$ \\

$f_n$-wide
& $\log\mathrm{U}(10^{-4},0.3)$
& $\mathrm{U}(40,90)$
& $\mathrm{U}(0.0058,0.026)$
& $1.35$ \\

$f_n$-narrow
& $\log\mathrm{U}(10^{-4},0.3)$
& $\mathrm{U}(52.5,55.5)$
& $\mathrm{U}(0.0058,0.026)$
& $3.71$ \\

$\beta$-wide
& $\log\mathrm{U}(10^{-4},0.3)$
& $\mathrm{U}(52,70)$
& $\mathrm{U}(0.003,0.04)$
& $1.49$ \\

$\beta$-narrow
& $\log\mathrm{U}(10^{-4},0.3)$
& $\mathrm{U}(52,70)$
& $\mathrm{U}(0.016,0.023)$
& $3.06$ \\

Joint-wide
& $\log\mathrm{U}(10^{-5},1.0)$
& $\mathrm{U}(40,90)$
& $\mathrm{U}(0.003,0.04)$
& $-0.57$ \\
\end{tabular}
\end{ruledtabular}
\end{table*}
\endgroup

\begin{figure*}[tbp]
\centering
\includegraphics[width=0.7\textwidth]{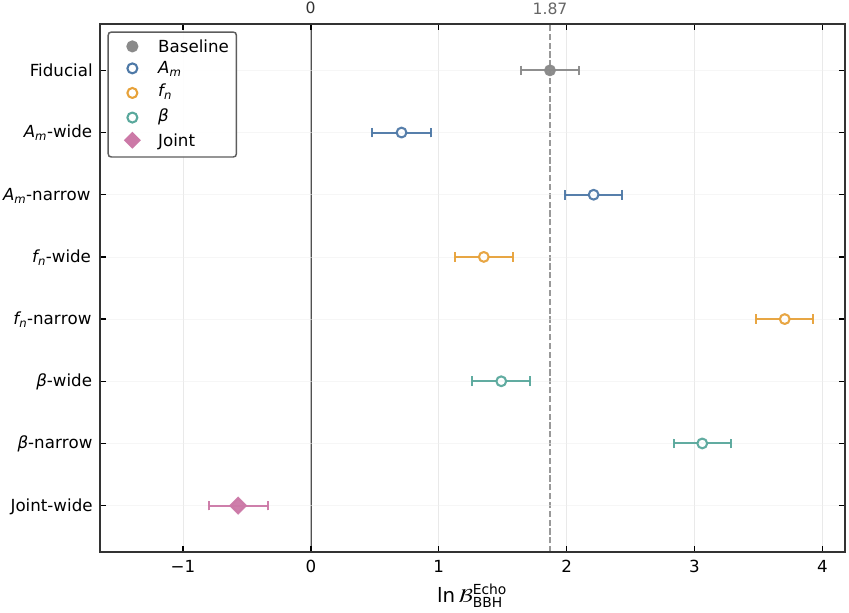}
\caption{\label{Echo_prior_sensitivity_plot}
Final values of $\ln \mathcal{B}^{\rm Echo}_{\rm BBH}$ for the prior configurations listed in Table~\ref{Echo_prior_sensitivity_table}. The solid 
and dashed vertical lines mark $\ln \mathcal{B}^{\rm Echo}_{\rm BBH}=0$ and the 
fiducial value $1.87$, respectively. Horizontal error bars show the uncertainties from the corresponding evidence calculations, and the diamond marker identifies the joint-wide configuration.}
\end{figure*}

Table~\ref{Echo_prior_sensitivity_table} summarizes the resulting values of $\ln \mathcal{B}^{\rm Echo}_{\rm BBH}$ for the tested prior configurations, which are shown in Fig.~\ref{Echo_prior_sensitivity_plot}. For the tested configurations, narrower prior intervals tend to increase $\ln \mathcal{B}^{\rm Echo}_{\rm BBH}$,
whereas wider intervals tend to decrease it. This behavior is consistent with the role of prior volume in the
evidence calculation: a narrower interval places more prior weight
on the region supported by the likelihood, whereas a wider
interval averages the likelihood over a larger parameter
volume~\cite{Lo:2018sep}. The joint-wide
configuration further reduces $\ln \mathcal{B}^{\rm Echo}_{\rm BBH}$ and gives
$\ln \mathcal{B}^{\rm Echo}_{\rm BBH}=-0.57$. Since the present
phenomenological model provides no physical prescription for selecting among
these prior ranges, the preference reported in the fiducial
analysis should therefore be regarded as a prior-dependent result rather than
a prior-independent conclusion.

\subsection{BBH-null calibration}
\label{subsec:bbh_null_calibration}

To quantify the impact of noise fluctuations on the fiducial
model-selection statistic
$T_{\rm obs}=\ln\mathcal{B}^{\rm Echo}_{\rm BBH}=1.87$, we construct
targeted BBH-null calibration ensembles by injecting a fixed GW231123-like
BBH signal generated with the \texttt{NRSur7dq4} waveform, with the
injection parameters listed in
Table~\ref{BBH_null_injection_parameters}, into simulated Gaussian noise and
off-source detector noise. Each resulting data realization is analyzed using the same Bayesian
setup as in the fiducial model comparison, with our phenomenological \texttt{sine-Gaussian} Echo waveform and the
\texttt{IMRPhenomXPHM-SpinTaylor} BBH waveform. The Gaussian-noise ensemble
contains 100 independent realizations, whereas the detector-noise ensemble
consists of 46 coincident H1--L1 off-source segments.\footnote{All calibration data sets are $4\,\mathrm{s}$ long and
sampled at $512\,\mathrm{Hz}$, matching the on-source data. They
are analyzed over $20$--$256\,\mathrm{Hz}$ using the same PSD
treatment and Bayesian analysis settings as in the fiducial
analysis.}

\begin{table}[t]
\caption{\label{BBH_null_injection_parameters}
Maximum-likelihood parameters for GW231123 inferred with
\texttt{NRSur7dq4} and used to generate the fixed GW231123-like
BBH injection for the targeted BBH-null calibrations. The
component masses are given in the detector frame.}
\centering
\begin{ruledtabular}
\begin{tabular}{lrlr}
Parameter & \multicolumn{1}{c}{Value}
          & Parameter & \multicolumn{1}{c}{Value} \\
\hline
$m_1\,(M_{\odot})$          & $156.83$
& $m_2\,(M_{\odot})$        & $143.59$ \\

$\chi_1$                       & $0.97$
& $\chi_2$                     & $0.92$ \\

$\theta_1\,(\mathrm{rad})$  & $1.32$
& $\theta_2\,(\mathrm{rad})$& $1.80$ \\

$\phi_{12}\,(\mathrm{rad})$ & $3.94$
& $\phi_{JL}\,(\mathrm{rad})$ & $4.81$ \\

$D_L\,(\mathrm{Mpc})$       & $1015.54$
& $\theta_{JN}\,(\mathrm{rad})$ & $1.74$ \\

$\alpha\,(\mathrm{rad})$    & $3.47$
& $\delta\,(\mathrm{rad})$  & $0.43$ \\

$\psi\,(\mathrm{rad})$      & $0.39$
& $\phi_c\,(\mathrm{rad})$  & $1.64$ \\
\end{tabular}
\end{ruledtabular}
\end{table}

For each calibration ensemble, we estimate the probability defined
in Eq.~\eqref{Eq:BBH_null_pvalue} using the finite-sample empirical
$p$-value estimator
\be
\hat{p}_\Lambda(T_{\rm obs}) =
\frac{N(T_i\geq T_{\rm obs})+1}{N_{\Lambda}+1},
\ee
where $T_i$ is the model-selection statistic computed for the
$i$th calibration realization, $N_{\Lambda}$ is the total number
of calibration realizations, and $N(T_i\geq T_{\rm obs})$ denotes
the number of realizations yielding a statistic greater than or
equal to the observed value. For the on-source value
$T_{\rm obs}=1.87$, the resulting empirical $p$-values are
$\hat{p}_{\rm G}\simeq0.18$ and
$\hat{p}_{\rm det}\simeq0.28$ for the simulated Gaussian-noise
and real off-source detector-noise ensembles, respectively.

\begin{figure*}[tbp]
\centering
\subfigure[]{
\includegraphics[width=0.65\textwidth]
{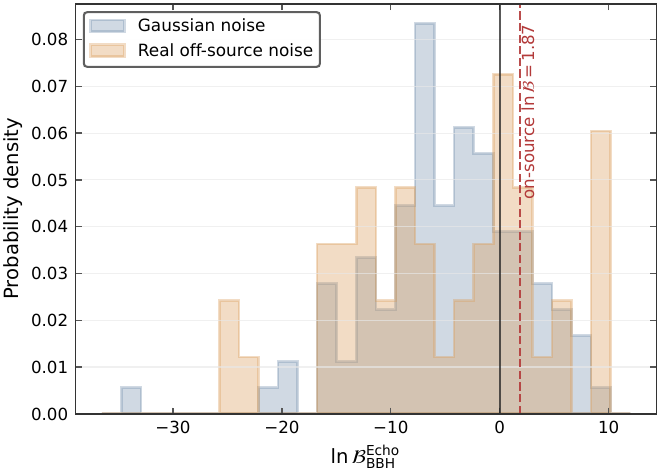}
}\\[0.6em]
\subfigure[]{
\includegraphics[width=0.65\textwidth]
{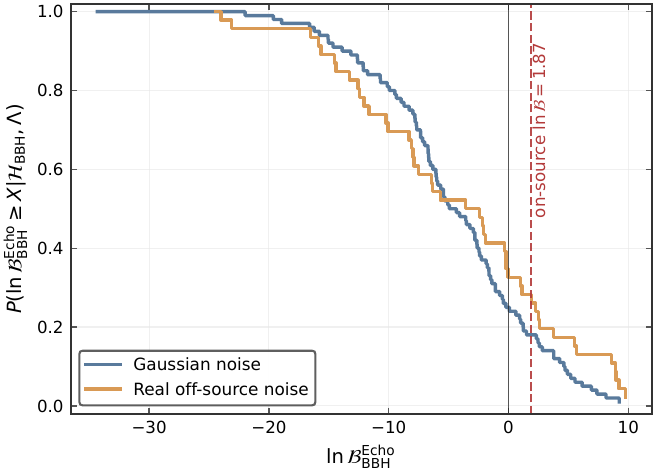}
}
\caption{\label{BBH_null_calibration_distributions}
Empirical calibration of the BBH-null model-selection statistic
$T=\ln\mathcal{B}^{\rm Echo}_{\rm BBH}$ using the targeted
BBH-null ensembles.
(a) Normalized histograms of $T$ obtained from the simulated
Gaussian-noise and real off-source detector-noise ensembles.
(b) Corresponding empirical complementary cumulative distribution
functions (CCDFs),
$P(T\geq X\mid\mathcal{H}_{\rm BBH},\Lambda)$,
where $\Lambda$ labels the calibration ensemble.
The solid vertical line marks equal support between the two
hypotheses, $T=0$, while the dashed vertical line marks the
on-source value, $T_{\rm obs}=1.87$.}
\end{figure*}

The empirical distributions of
$T=\ln\mathcal{B}^{\rm Echo}_{\rm BBH}$ are shown in
Fig.~\ref{BBH_null_calibration_distributions}. The detector-noise ensemble
exhibits a broader distribution and a more prominent high-$T$ tail than the
Gaussian-noise ensemble. For GW231123-like short-duration signals,
noise-induced distortions can appreciably affect the relative evidence for
competing models. In particular, realistic detector noise may contain
transient or other non-Gaussian features that are not represented in the
Gaussian-noise ensemble~\cite{Hu:2025lhv,Powell:2018csz,Kwok:2021zny}. Such features may be partially accommodated by the
phenomenological Echo model, which can describe localized oscillatory power
without enforcing the full BBH inspiral-merger-ringdown dynamics, thereby
increasing the probability of obtaining a positive
$\ln\mathcal{B}^{\rm Echo}_{\rm BBH}$ under the BBH hypothesis.

\section{Conclusions and outlook}
\label{sec:conclusion}

In this work, we applied a Bayesian model-comparison analysis to
the short-duration event GW231123, comparing the standard BBH
interpretation with a phenomenological echo-for-wormhole
alternative. We modeled the putative leading Echo pulse with a
\texttt{sine-Gaussian} wavepacket and adopted
\texttt{IMRPhenomXPHM-SpinTaylor} as the fiducial BBH waveform.
The two models recover comparable network SNRs, while the
fiducial Bayesian comparison gives
$\ln\mathcal{B}^{\rm Echo}_{\rm BBH}=1.87$.
In the same overall framework, our previous analysis of GW190521
yielded
$\ln\mathcal{B}^{\rm Echo}_{\rm BBH}\approx-2.9$
\cite{Lai:2025skp}. Thus, under the respective analysis setups, GW231123 yields a more Echo-favoring value of the
model-selection statistic than GW190521. This comparison, however, remains
conditional on the adopted waveform implementations and priors.

We further assessed the robustness of the GW231123 result against
BBH waveform systematics, Echo-prior choices, and noise
fluctuations. All three alternative BBH waveform models yield
negative values of
$\ln\mathcal{B}^{\rm Echo}_{\rm BBH}$, ranging from $-8.01$ to
$-3.64$, while broader Echo priors generally reduce the Bayes
factor, with the joint-wide configuration giving
$\ln\mathcal{B}^{\rm Echo}_{\rm BBH}=-0.57$. The targeted BBH-null
calibrations yield empirical
$p$-values of
$\hat p_{\rm G}\simeq0.18$ and
$\hat p_{\rm det}\simeq0.28$. These values indicate that, for
GW231123-like BBH signals in the tested noise ensembles, an
apparent Echo preference at least as strong as that observed is
not statistically rare. The fiducial positive Bayes factor is
therefore better regarded as a configuration-dependent result
than as a robust feature of the data. Overall, these results favor the
BBH interpretation of GW231123 under the analysis configurations
considered here.

Establishing whether this conclusion remains valid beyond the
model space explored here will require advances in both waveform
modeling and empirical calibration. On the modeling side, future
investigations should develop more physically complete
wormhole-echo waveforms that go beyond the present single-pulse
phenomenological description and incorporate additional physical
effects, including spin-dependent modifications,
frequency-dependent reflectivity, backreaction, and more general
multi-pulse structures
\cite{Cardoso:2016oxy,Bueno:2017hyj,Xin:2021zir,
Oshita:2019sat,Vellucci:2022hpl,Wang:2019szm}%
\footnote{It is worth noting that mergers of supermassive BBHs,
the most massive coalescence events, may provide a plausible
setting for the formation of a transient wormhole. Such scenarios
may also be relevant to the origin of our Universe,
e.g.,~\cite{Piao:2003zm,Piao:2005ag,Liu:2011ns,Lan:2024gnv}.
Recently, supermassive primordial black holes have also been
considered as possible sources of the high-redshift massive
galaxies observed by JWST
\cite{Huang:2023chx,Huang:2023mwy,Huang:2024aog,
Huang:2024koy,Hai-LongHuang:2024gtx,Li:2025jxq}.}.
On the statistical side, more extensive empirical calibrations
using real detector noise, multiple BBH waveform families, and
injections spanning the relevant source population will be
important for quantifying the significance of apparent model
preferences. Template-based Bayesian analyses can also be
complemented by morphology-independent or model-independent
searches
\cite{Tsang:2018uie,Wu:2023wfv}. As the sample of massive,
short-duration GW events grows, applying these methods within a
uniform analysis framework may help determine whether similar
features recur across multiple events and provide more stringent
tests of alternative post-merger interpretations.

\textbf{Acknowledgments}

This work is supported by National Key Research and Development
Program of China, No.2021YFC2203004, and NSFC, No.12475064, and
the Fundamental Research Funds for the Central Universities.

\end{document}